\documentclass[12pt,a4paper]{conference}

\usepackage{fancyhdr}
\usepackage{graphicx,amsmath,amssymb,cite}
\usepackage{multind}
\makeindex{author} \makeindex{subject}
\newcommand{\beq}{\begin{equation}}
\newcommand{\eeq}[1]{\label{#1}\end{equation}}
\newcommand{\eeqn}{\end{equation}}

\newcommand{\beqa}{\begin{eqnarray}}
\newcommand{\eeqa}[1]{\label{#1}\end{eqnarray}}
\newcommand{\eeqan}{\end{eqnarray}}

\let\bar=\overbar

\newcommand{\Dslash}{\not{\hbox{\kern-4pt $D$}}}
\newcommand{\dslash}{\not{\hbox{\kern-2pt $\del$}}}

\newcommand{\msb}{{\bar{\ssstyle M \kern -1pt S}}}

\begin{document}

\Chapter{Exotic hadrons \\
and SU(3) chiral dynamics
}
           {Exotic hadrons and SU(3) chiral dynamics}{T. Hyodo \it{et al.}}
\vspace{-6 cm}\includegraphics[width=6 cm]{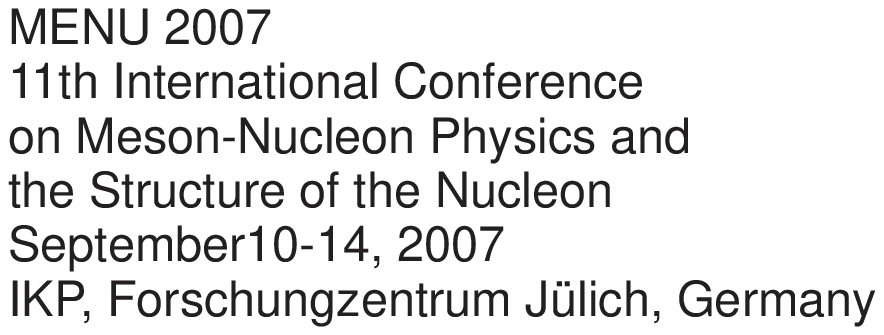}
\vspace{4 cm}

\addcontentsline{toc}{chapter}{{\it T. Hyodo}} \label{authorStart}

\begin{raggedright}


Tetsuo Hyodo$^{\star,\%}$$^,$\footnote{E-mail address:
thyodo@ph.tum.de}~, 
Daisuke Jido$^{\%}$,
Atsushi Hosaka$^{\#}$ 
\bigskip\bigskip


$^{\star}$Physik-Department, Technische Universit\"at M\"unchen, 
D-85747 Garching, Germany\\
$^{\%}$Yukawa Institute for Theoretical Physics (YITP),
Kyoto University, Kyoto, 606-8502, Japan\\
$^{\#}$Research Center for Nuclear Physics (RCNP),
Ibaraki, 567-0047 Japan\\

\end{raggedright}


\begin{center}
\textbf{Abstract}
\end{center}
We explore a possibility to generate exotic hadrons dynamically in the 
scattering of hadrons. The $s$-wave scattering amplitude of an arbitrary 
hadron with the Nambu-Goldstone boson is constructed so as to satisfy the 
unitarity condition and the chiral low energy theorem. We find that the 
chiral interaction for the exotic channels is in most cases repulsive, and 
that the strength of the possible attractive interaction is uniquely 
determined. We show that the attractive interaction in exotic channels is not
strong enough to generate a bound state, while the interaction in nonexotic 
channel generate bound states which are considered to be the origin of some 
resonances observed in nature.

\vspace{1 cm}


Strong interaction of QCD exhibits rich spectra of hadrons in the 
nonperturvative vacuum at low energy, where about 300 hadronic states have 
been observed~\cite{Yao:2006px}. It is important to investigate the 
properties of hadrons to understand the low energy dynamics of QCD. Chiral 
symmetry provides us a way to study hadron properties in connection with the 
fundamental theory of QCD.

Dynamical models based on chiral symmetry, known as chiral unitary approach,
successfully describe the two-body scattering of hadrons with the 
Nambu-Goldstone (NG) bosons in coupled channels, dynamically generating some 
$s$-wave resonances in the scattering~\cite{Kaiser:1995eg,Oset:1998it,
Oller:2000fj,Lutz:2001yb}. These studies are along the same line with the 
coupled-channel dynamical models for the meson-baryon scattering studied in 
60's, where the vector meson exchange interaction was adopted. This 
phenomenological interaction is now identified as the Weinberg-Tomozawa (WT) 
term, which is the leading order term 
in chiral perturbation theory. In this respect, one can introduce higher 
order corrections into the interaction systematically. The WT interaction was
originally derived in current algebra. Since current algebra tells us about 
the interaction for arbitrary target hadrons, it is possible to apply the 
chiral unitary approach to the system with spin $3/2$ baryons with the 
decuplet baryons as target, to the heavy quark sectors, and to the axial 
vector mesons. In the series of studies, the properties of the generated 
resonances are in fair agreement with experimental data.

On the other hand, the hadrons observed so far can be classified by their 
flavor quantum numbers. \textit{Empirically}, there is a regularity in the 
quantum numbers of the observed hadrons: the states with the valence quark 
contents of $\bar{q}q$ or $qqq$ were observed, while no state was well
established with larger number of valence quarks (4, 5, 6,\ldots quarks). The
latter states, called exotic hadrons, were intensively studied recently after
the report on the $\Theta^+$ by LEPS collaboration~\cite{Nakano:2003qx}. In 
spite of the large amount of theoretical works, it is not clear why the 
exotic hadrons are difficult to observe.

In order to clarify this issue, we have recently performed an analysis of 
exotic hadrons in $s$-wave chiral dynamics~\cite{Hyodo:2006yk,Hyodo:2006kg,
Hyodo:2006sw,Hyodo:2007jk}. We utilize the framework of the chiral unitary 
approach, since it is naively expected that the resonances produced in the 
dynamical model should have large component of the multiquark configuration, 
which is the flavor partner of the exotic hadrons in the $s$-wave 
scattering. 

We construct the scattering amplitude of an arbitrary hadron with the 
Nambu-Goldstone boson $t(\sqrt{s})$ as
\begin{align}
    t(\sqrt{s}) &\to V^{\text{chiral}}(\sqrt{s}) \quad \text{at low energy}
    \label{eq:chiral} , \\
    \text{Im} t^{-1}(\sqrt{s}) & =\frac{\rho(\sqrt{s})}{2}
    \label{eq:unitary} ,
\end{align}
where $V^{\text{chiral}}(\sqrt{s})$ is the low energy interaction based on 
chiral symmetry and $\rho(\sqrt{s})$ is the phase space of the two-body 
scattering. Eq.~\eqref{eq:chiral} is the constraint from the chiral low 
energy theorem, whereas Eq.~\eqref{eq:unitary} guarantees the unitarity of 
the S-matrix. Utilizing this approach, we would like to study what chiral 
dynamics tells us about the existence of the exotic hadrons.

The low energy $s$-wave interaction of a target hadron $(T)$ with the NG 
boson in a channel $\alpha$ is given by
\begin{equation}
    V_{\alpha }
    =-\frac{ \omega}{2f^2}C_{\alpha,T}  , 
    \label{eq:WTint}
\end{equation}
where $\omega$ and $f$ are the energy and the decay constant of the NG boson,
and the expression for the group theoretical factor $C_{\alpha,T}$ is given 
in Refs.~\cite{Hyodo:2006yk,Hyodo:2006kg,Hyodo:2006sw,Hyodo:2007jk}. By 
examining the coupling strength $C_{\alpha,T}$ for exotic channels, we find 
that the interaction for exotic channels is in most cases repulsive, and the
strength of the possible attractive interaction is uniquely determined as
\begin{equation}
    C_{\text{exotic}}=1  . \label{eq:Exoticattraction}
\end{equation}
Eqs.~\eqref{eq:WTint} and \eqref{eq:Exoticattraction} determines the low 
energy interaction $V_{\text{chiral}}$ in Eq.~\eqref{eq:chiral} for exotic 
channels.

Next we construct the scattering amplitude consistent with 
Eq.~\eqref{eq:unitary}. Based on the $N/D$ method\cite{Oller:2000fj}, the 
general form of the scattering amplitude $t_{\alpha}(\sqrt{s})$ can be 
written down with one subtraction constant, which is determined by the 
requirement~\eqref{eq:chiral} at the scale 
$\sqrt{s}=M_T$~\cite{Hyodo:2006kg}. Thus we obtain the scattering amplitude 
$t_{\alpha}(\sqrt{s})$ which satisfies both Eqs.~\eqref{eq:chiral} and 
\eqref{eq:unitary}. We then search for poles of bound states in the amplitude
$t_{\alpha}(\sqrt{s})$. From the energy dependence of the interaction and the
loop function, we find the critical value for the attractive interaction 
strength which is enough to make a bound state as
\begin{align}
    C_{\text{crit}}= \frac{2f^2 }{m\bigl[-G(M_{T}+m)\bigr]} ,
    \label{eq:WTcritical}
\end{align}
where $G(\sqrt{s})$ is defined as once-subtracted dispersion 
integral~\cite{Hyodo:2006kg}. If the interaction strength $C_{\alpha,T}$ is 
larger than this critical value, a bound state is generated in the amplitude.
Comparing $C_{\text{crit}}$ with the attractive interaction in the exotic 
channel~\eqref{eq:Exoticattraction}, we show that the interaction is not 
strong enough to generate a bound state for the mass of the target hadron 
smaller than 6 GeV.

In this way, we have studied the exotic states in the NG boson-hadron 
scattering. We construct the scattering amplitude which satisfies the chiral
low energy theorem and unitarity condition. Considering an arbitrary target 
hadron, we find that the interaction in the exotic channels are in most 
cases repulsive, and possible attractive interaction is uniquely given as 
$C_{\rm exotic}=1$. We show that the strength of the attractive interaction 
is not sufficient to generate a bound state for the physically known masses 
of the target hadrons.

In order to draw a general and model-independent conclusion, we have 
simplified the framework of the chiral unitary approach. Our basic 
assumptions are 1) flavor SU(3) symmetry and 2) dominance of the 
leading order interaction. Once we accept these conditions, the subsequent 
arguments are straightforward. In practice, however, the SU(3) symmetry is 
broken and the higher order terms of the chiral expansion would play a 
substantial role, especially for the NG boson with larger masses. These 
effects could be included in the kernel interaction based on chiral 
perturbation theory, but we need experimental data to determine the low 
energy constants. It should be noted that the exotic hadrons constructed by 
other mechanisms than the present analysis, such as by quark dynamics, are 
not excluded.

In this study, we stress that the WT term is the leading order term of the
chiral expansion and the strength is only determined by the group theoretical
factor. We can therefore argue that the leading order term does not provide 
bound states in exotic channel, without performing experiments. Given the 
success of the chiral unitary approach in the nonexotic sectors which our 
arguments are based on, our result may partly explain the difficulty to 
observe exotic hadrons in nature.

\section*{Acknowledgements}
T.H. 
appreciates Prof. W. Weise for the discussion about the convergence of chiral
expansion. T.~H. thanks the Japan Society for the Promotion of Science (JSPS)
for financial support.  This work is supported in part by the Grant for 
Scientific Research (No.\ 17959600, No.\ 19853500, No.\ 18042001, and 
No.\ 16540252) and by Grant-in-Aid for the 21st Century COE "Center for 
Diversity and Universality in Physics" from the Ministry of Education, 
Culture, Sports, Science and Technology (MEXT) of Japan.

%
\end{document}